\documentclass{article}
\usepackage{RR}
\usepackage{boxedminipage, graphicx, float, amsmath, amssymb, amsxtra, latexsym}
\usepackage{picinpar, wrapfig}
\input epsf

\newfloat{Table}{hbt}{mytab}[section]
\newcommand{\block}[3]{#1 \ \ \left\{\ \begin{array}{l} #2 \\ #3 \\ \end{array}\right.}
\RRdate{Mai 2006}
\RRauthor{Laurent Hasco\"et \and Mauricio Araya-Polo}
\authorhead{Hasco\"et and Araya-Polo}
\RRetitle{Enabling user-driven Checkpointing strategies in Reverse-mode Automatic Differentiation}
\RRtitle{Strategies de checkpointing pilotees par l'utilisateur en Differentiation Automatique inverse}
\titlehead{User-driven Checkpointing strategies}
\RRresume{Nous pr\'esentons une nouvelle fonctionnalit\'e de l'outil de
Diff\'erentiation Automatique (DA) \textsc{tapenade}.
Le mode inverse de la DA construit des codes adjoints, qui sont
largement utilis\'es en calcul scientifique, pour l'optimisation ou les probl\`emes
inverses. Bien qu'a priori remarquablement efficace, le mode inverse
souffre de la tr\`es grande consommation m\'emoire requise pour
conserver des valeurs interm\'ediaires du programme initial.
Le Checkpointing est un compromis stockage-recalcul qui r\'eduit
cette consommation. Notre but est de rendre l'utilisation du Checkpointing
dans \textsc{tapenade} plus flexible, en particulier par des directives utilisateur.
Notre but \`a terme est de d\'evelopper des strat\'egies
semi-automatiques optimales d'application du Checkpointing.
Dans ce rapport, nous pr\'esentons les modifications apport\'ees \`a \textsc{tapenade}
pour rendre le Checkpointing flexible, puis nous \'etudions et nous comparons certaines
strat\'egies de Checkpointing sur plusieurs applications r\'eelles
provenant d'utilisations industrielles de \textsc{tapenade},
dans le but de d\'egager des heuristiques g\'enerales.
Certaines exp\'eriences montrent des am\'eliorations en m\'emoire de l'ordre de 35\%,
et en temps d'ex\'ecution de l'ordre de 90\%.}
\RRabstract{This paper presents a new functionality of the Automatic Differentiation (AD)
Tool \textsc{tapenade}. \textsc{tapenade} generates adjoint codes which 
are widely used for optimization or inverse problems. Unfortunately, for large
applications the adjoint code demands a great deal of memory, because it needs to store a 
large set of intermediates values. To cope with that problem,
\textsc{tapenade} implements a sub-optimal version of a technique
called checkpointing, which is a trade-off between storage and recomputation.
Our long-term goal is to provide an optimal checkpointing 
strategy for every code, not yet achieved by any AD tool. Towards that goal, we first 
introduce modifications in \textsc{tapenade} in order to give the user the choice to 
select the checkpointing strategy most suitable for their code. Second, we conduct 
experiments in real-size scientific codes in order to gather hints that help us to 
deduce an optimal checkpointing strategy. Some of the experimental results show 
memory savings up to 35\% and execution time up to 90\%.}
\RRmotcle{Diff\'erentiation Automatique, Mode Inverse, Checkpointing, TAPENADE}
\RRkeyword{Automatic Differentiation, Reverse Mode, Checkpointing, TAPENADE}
\RRprojet{Tropics}
\RRtheme{\THNum}
\URSophia

\begin{document}
\makeRR

\section{INTRODUCTION}\label{sec:intro}

The context of this work is Automatic Differentiation (AD)~\cite{Corliss01,Griewank00}. 
The reverse mode of AD is a promising way to build adjoint codes to compute gradients.
The fundamental advantage of adjoint codes is that they compute gradients
at a cost which is independent of the dimension of the input space, and they are thus
a key ingredient to solve inverse problems and optimization
problems~\cite{Jameson,Courty}.
AD adjoint codes are fundamentally made of two succe\-ssi\-ve sweeps,
a {\em forward} sweep running the original code and storing a 
significant part of the intermediate values, and a {\em backward} sweep using these 
values to compute the derivatives. For large applications, such as CFD programs,
reverse differentiated codes may end up using far too much memory.\\

Checkpointing is a standard time/memory trade-off tactic to reduce the peak of this 
memory use. When a segment of the program is checkpointed, it is executed {\em without} 
storage of the intermediate values. Later on, when the backward sweep reaches the 
checkpointed segment, this segment must be executed a second time {\em with} storage, 
and finally the backward sweep may resume. 
Checkpointing has a benefit: there are two places where the memory size
reaches a peak, namely at the end of the forward sweep and at the end of the
checkpointed segment, and both peaks are generally smaller than the peak without
checkpointing. On the other hand, checkpointing has a cost: {\it (1)} in execution time 
because segment is executed twice and {\it (2)} in memory because intermediate values
must be store to run the segment twice. Hopefully this last me\-mo\-ry cost is less
than the memory benefit above.\\

In AD tools, checkpointing is applied systematically, for instance at procedure 
calls or around loops bodies. Experience shows that checkpointing every procedure 
call is in general sub-optimal. Optimal strategies have been found only for 
the case of a fixed-length loop~\cite{Griewank92}, and not for the nested 
procedure structure of real-life codes.\\

Towards the ultimate goal of an AD tool embedding an optimal checkpointing strategy 
for all programs, we propose in a first step to activate checkpointing for only a 
number of user-selected procedure calls. Therefore, in addition to the default 
systematic checkpoint mode (called {\em joint} mode in~\cite{Griewank00}), each 
procedure may now be differentiated in the so-called {\em split} mode, i.e. without
checkpointing. In split mode, the procedure gives rise to two separate differentiated 
procedures, one for the forward sweep and one for the backward sweep.\\

This paper presents the implementation of this new split mode functionality inside 
our AD tool \textsc{tapenade}~\cite{Hasco04}, which up to now only featured the joint mode. We 
also discuss the necessary adaption of the existing preliminary data-flow analyses 
namely, adjoint-liveness analysis~\cite{Hasco05} and TBR analysis~\cite{Hasco02, Hasco05}.
In a second step, we use this user control on checkpointing to make experimental 
measurements of various checkpointing choices on several large scientific codes. 
We present the results of these experiments, some of which show savings of memory 
up to 35\% and execution time up to 90\%. Also, these results give hints to a general 
automatic strategy of where to use checkpointing. At present, no AD tool
has such a general checkpointing strategy, and our long term goal is to provide
one in \textsc{tapenade}.\\

The remainder of this paper is structured as follows: Section~\ref{sec:ad} introduces
the reverse mode of AD. In Section~\ref{sec:check} we present the checkpointing 
technique and show how different checkpointing placement strategies affect the
behavior of the reverse differentiated code. In Section~\ref{sec:imple} we discuss 
the implementation issues. In Section~\ref{sec:expe} we present and discuss the experimental 
measurements. Finally, we discuss the future work and the conclusions in Section~\ref{sec:conclu}.
\section{REVERSE AUTOMATIC DIFFERENTIATION}\label{sec:ad}
In our context, AD is a program transformation technique to obtain derivatives,
and in particular gradients. We are given a program $P$ that evaluates a function
$F$. Program $P$ can be seen as a sequential list of instructions $I_{j}$
$$P = I_{1}\ ;\ I_{2}\ ;\ \ldots\ ;\ I_{j}\ ; \ldots\ ;\ I_{p-1}\ ;\ I_{p},$$
where the instructions represent elementary functions $f_{i}$.
Then the function $F$ is indeed
$$F = f_{p} \circ f_{p-1} \circ \ldots  \circ f_{j} \circ \ldots \circ f_{2} \circ f_{1}.$$
AD takes advantage of this to apply the chain rule of calculus to build a
new program that evaluates the derivatives of $F.$\\
The {\it reverse} mode of AD computes gradients. Roughly speaking, for a given scalar output, it 
returns the direction in the input space that maximizes the increase of this output.
Strictly speaking gradient is defined only for scalar output functions. Therefore, 
we build a vector $\overline Y$ that defines the weights of each component of the 
original output $Y = F(X)$. This defines a scalar output $\overline{Y}^{t}\times Y = Y^{t}
\times \overline{Y} = F^{t}(X)\times \overline{Y}$. Its gradient 
has thus the following form:
\begin{equation}\label{form:reverse1}
\overline{X} = F'^{t}(X) \times \overline{Y} = f'^{t}_{1}(x_{0}) \times \ldots \times f'^{t}_{j+1}(x_{j}) \times \ldots \times f'^{t}_{p}(x_{p-1})\times \overline{Y} 
\end{equation}
where $x_{i-1}$ is the set of all variables values just before execution of 
the instruction that implements $f'^{t}_{i}$, and $F'^{t}(X)$ is the transposed
Jacobian.\\

Formula~\ref{form:reverse1} is implemented from right to left, because matrix$\times$vector 
products are cheaper to compute than matrix$\times$matrix products. This result 
in probably the most efficient way to compute a gradient. Unfortunately, this mode of 
AD has a difficulty: the $f'^{t}_{i}$ instructions require the in\-ter\-me\-diate
values $x_{i-1}$ in the reverse of their creation order. The trouble is that
programs often overwrite variables, and therefore these values may be lost when 
needed by the $f'^{t}_{i}$. \\

\begin{figure}[!ht]
\centering
\includegraphics[height=2.8in, width=4.8in]{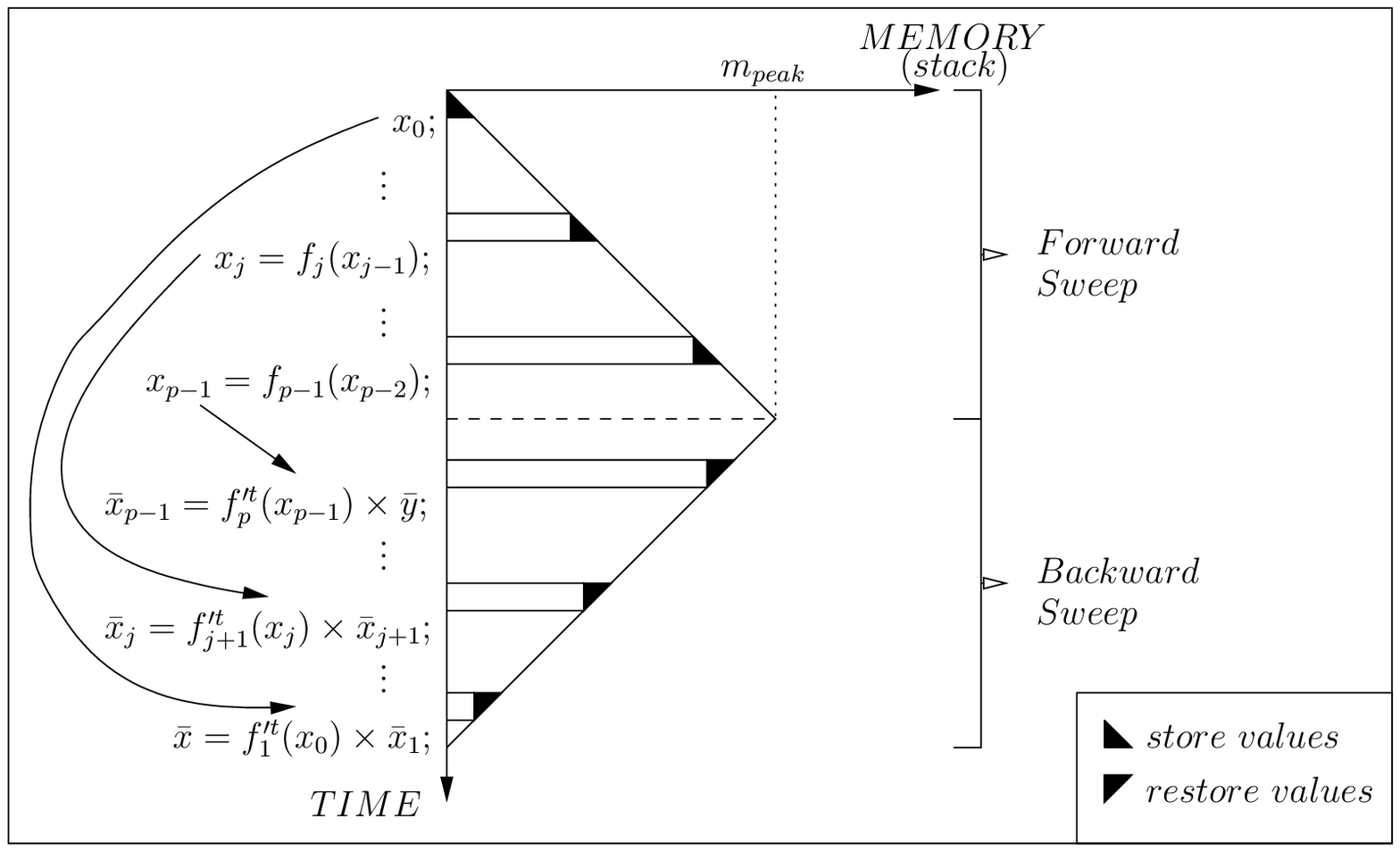}
\caption{\label{fig:tape} The horizontal axis represents the amount of values
currently on the stack.}
\end{figure}

There are two main strategies to cope with this problem:
{\it Recompute-All}~\cite{Giering97} or {\it Store-All}~\cite{Griewank00}. Recompute-All strategy is 
very demanding in execution time, quadratic with respect to the number of 
run-time instructions, because it recomputes the intermediates values every time they are 
required, from a saved initial point.
On the other hand, the Store-All strategy is linear with respect to
the number of run-time instructions, both for
memory consumption and execution time, because it consists in storing on a stack
all values required later by derivatives, and then restore them
when they are needed. This results in the structure of reverse 
differentiated programs shown on Figure~\ref{fig:tape}.\\

Because we will need to reason formally about adjoint programs in the sequel of
this paper, we need to denote them in a more algebraic way.  The 
reverse differentiated program $\overline P$ has two parts. The first is called the 
forward sweep $\overrightarrow P$, and is basically the necessary ``slice" of 
the original program $P$ plus some instructions to store required va\-lues. 
The second part is called the backward sweep $\overleftarrow P$, and consists 
of the instructions that implement the functions $f'^{t}_{i}(x)$
from Formula~\ref{form:reverse1}, plus some instructions to recover the needed
intermediate values. \\

Formalizing the structure of the program in Figure~\ref{fig:tape},
the structure of the reverse differentiated program $\overline{P}$ of a program $P$
is roughly described by equation~(\ref{form:reverse2})
\begin{equation}\label{form:reverse2}
\overline{P} = \overrightarrow P\ ;\ \overleftarrow P = I_{1}\ ; \dots ;\ I_{p-1}\
;\ \overleftarrow{{I}_{p}}\ ; \dots;\ \overleftarrow{I_{1}}
\end{equation}

Figure~\ref{fig:example} shows the reverse differentiated version of a small
example procedure, featuring the forward and backward sweeps.
The {\tt PUSH()} and {\tt POP()} calls
store and restore values of required intermediates variables. We can now
refine formula~(\ref{form:reverse2}) by inserting these calls.
For any instruction $I$ and any program 
tail $D$ after $I$, the program $\overline{P}$ is defined
recursively by the following equation:
\begin{equation}\label{form:reverse3}
\overline{P} = \overline{I\ ;\ D} = \overrightarrow{I}\ ;\ \overline{D}\ ;\ \overleftarrow{I} = {\tt
PUSH}(\hbox{\bf out}(I))\ ;\ I\ ;\ \overline{D}\ ;\ {\tt POP}(\hbox{\bf out}(I))\ ;\ I'
\end{equation}
where $\hbox{\bf out}(I)$ is a set of values overwritten by instruction $I$.
In reality, we store only the intermediates values which are required to compute
the derivatives of $I$ and of its preceding instructions.
The data-flow equations of the static analysis that evaluates these
values "To Be Recorded", known as the "TBR" analysis,
was given in~\cite{Hasco05}.
\begin{figure}[!ht]
\begin{tabular}{|l@{\hspace{1.6in}}|l@{\hspace{2.7in}}|}
\hline
\multicolumn{1}{|c|}{\bf Original procedure}&\multicolumn{1}{c|}{\bf Reverse differentiated procedure}\\ \hline\hline
\begin{minipage}[t][1.0in][t]{1.0in}
{\tt
\begin{tabbing}
- \= ---- \= ---- \= ... \kill
\>\>subroutine sub1(x,y,z)\>\\
\>\>\\
\>$I_{1}$\>   tmp1 = SIN(y)\>\\
\> \>\\
\>$I_{2}$\>   y = y * y\>\\
\> \>\\
\>$I_{3}$\>   tmp1 = tmp1 * x\>\\
\>$I_{4}$\>   z = y / tmp1\>\\
\>\> end\>\\
\end{tabbing}}
\end{minipage}
&
\begin{minipage}[t][3.4in][t]{1.0in}
{\tt
\begin{tabbing}
- \= ------ \= --- \= ... \kill
\>\>  subroutine sub1\_b(x,xb,y,yb,z,zb)\>\\
\> \>\\
\>$I_{1}$\>   tmp1 = SIN(y)\>\\
\>\>   PUSH(y)\>\\
\>$I_{2}$\>   y = y * y\>\\
\>\>   PUSH(tmp1)\>\\
\>$I_{3}$\>   tmp1 = tmp1 * x\>\\
<{\em forward sweep ends, backward sweep begins}>\>\>\\
\>$\block{I'_{4}}{{\tt yb = zb / tmp1}}{{\tt tmp1b = -(y * zb / tmp1 * * 2)}}$\>\>\\
\>\>   POP(tmp1)\>\\
\>$\block{I'_{3}}{{\tt xb = tmp1 * tmp1b}}{{\tt tmp1b = x*tmp1b}}$\>\>\\
\>\>   POP(y)\>\\
\>${I'_{2}}$\> yb = 2 * y * yb\>\\
\>${I'_{1}}$\> yb = COS(y) * tmp1b\>\\
\>\>  end\>\\
\end{tabbing}}
\end{minipage}\\ \hline
\end{tabular}
\caption{\label{fig:example} The structure of a reverse differentiated program}
\end{figure}
\newpage
\section{CHECKPOINTING}\label{sec:check}
To control the memory problem caused by the storage of intermediates values, 
the Store-All strategy can be improved in two main directions: {\it (1)} refine the data-flow 
analyses in order to reduce the number of values to store, and {\it (2)} deactivate the 
Store-All strategy for chosen segments of the code, therefore saving memory space. 
The former is described in~\cite{Hasco05,Hasco06}, the latter is the focus of this
work.\\ 

The mechanism which deactivates the Store-All strategy for certain chosen segment
is called {\em checkpointing}. It has two consequences on the behavior of the
reverse differentiated program:
\begin{enumerate}
\item when the backward sweep reaches the chosen segment, it must be executed again, 
this time with Store-All strategy turned on.
\item in order to execute the segment twice, a sufficient set of values (called
a {\it snapshot}) must be stored before the first execution of the segment.
\end{enumerate}
\begin{figure}[!ht]
\centering
\includegraphics[height=6.0in, width=3.8in]{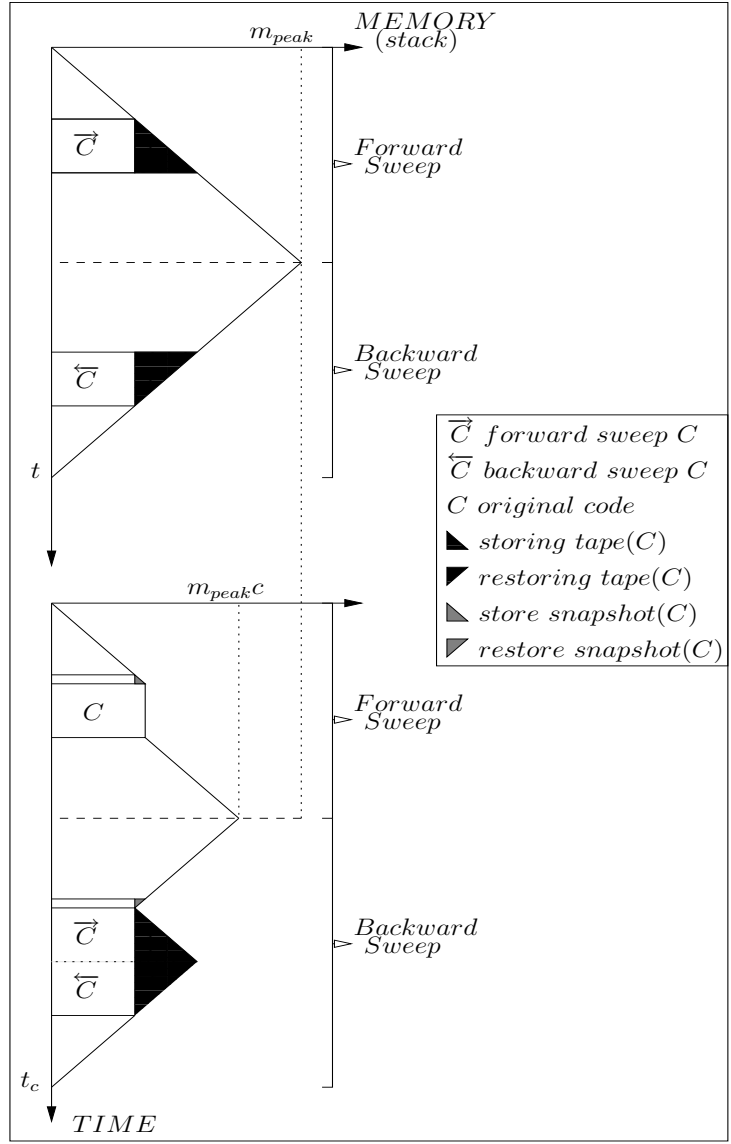}
\caption{\label{fig:tape-check} Checkpointing in Reverse Mode AD.}
\end{figure}

On Figure~\ref{fig:tape-check}, we assume that $snapshot(C) < tape(C)$.  This
is a reasonable assumption in most cases, and particularly when C is large. As
a consequence we see that $m_{peak}c$ is smaller than $m_{peak}$, because in 
the checkpointed case the first execution of segment C does not store anything.
Conversely, we see that the time $t_{c}$ is longer than $t$, because in
the no-checkpointing case every piece of the code is executed only once, whereas
we observe in the checkpointing case that segment $C$ is executed
twice ($C$ and $\overrightarrow{C}$).\\   

Checkpointed segments can be chosen in different ways, and can be nested.
One classical strategy is to checkpoint each and every procedure call.
However, experience indicates that this strategy is not optimal, though the 
optimal situation is not easy to foresee. 
Since the optimal checkpointing strategy is still out reach, it seems natural to 
let the user influence the choice. A completely user-driven checkpointing
will allow the user to try each and every combination, looking for an optimal
placement of checkpoints.
This paper describes the developments to achieve this user interaction.
In a second step, this will let us experiment about rules and tactics, towards the 
long-term goal of computer aided optimal checkpointing.
This paper presents our first experiments in this direction. \\

The assumption behind checkpointing is that $snapshot(C) < tape(C)$. To keep
the snapshots small, we need to develop the algebraic notation of
equation~(\ref{form:reverse3}). When segment $C$ is checkpointed
(denoted with surrounding parentheses), reverse differentiation of the
program $P=U;C;D$ is defined by the recursive equation

\begin{equation}\label{form:call2}
\overline{P} = \overline{U;(C);D} = \overrightarrow{U}; {\tt PUSH}(\hbox{\bf snp}(C));
C; \overline{D}; {\tt POP}(\hbox{\bf snp}(C)); \overline{C}; \overleftarrow{U}
\end{equation}

where $U$/$D$ are the code segments $U$pstream/$D$ownstream of $C$
and $\hbox{\bf snp}(C)$ is the snapshot stored to re-execute $C$.
Intuitively, if a variable is not modified by $C$ nor by $\overline{D}$, then
its value will be unmodified when $C$ is run again and it is not necessary to store it.
We shall denote by {\bf out}$(X)$ the set of variables overwritten by the code segment $X$.
Also, only the variables that are going to be used by $C$ need to be in the snapshot.
Indeed, only the variables that are used by $\overline{C}$ need to be stored, and
this set is often smaller than the variables used by $C$. We shall call it
$\hbox{\bf live}(\overline{C})$, and it is determined by the so-called
{\em adjoint liveness analysis}.
Therefore a good enough conservative definition of the snapshot is:
\begin{equation}\label{form:snap}
\hbox{\bf snp}(C) = \hbox{\bf live}(\overline{C}) \cap (\hbox{\bf out}(C) \cup \hbox{\bf out}(\overline{D}))
\end{equation}
 The data-flow equations of adjoint liveness analysis were
defined formally in~\cite{Hasco05}.
Snapshots can be refined further, taking into account the interactions between
successive or nested checkpointed segments. A study on minimal snapshots can be
found in~\cite{Hasco06}.

Let's now focus on the checkpoint placement problem.
In \textsc{tapenade} like in many other AD tools, the natural checkpointed segment
is the procedure call. Therefore in the sequel we shall experiment with
various placements of checkpoints, all around procedure calls, and therefore
shown on call trees. This hypothesis is by no means restrictive and our conclusions
can be extended to arbitrary cleanly nested code segments.
\begin{figure}[!ht]
\centering
\includegraphics[height=1.6in, width=4.7in]{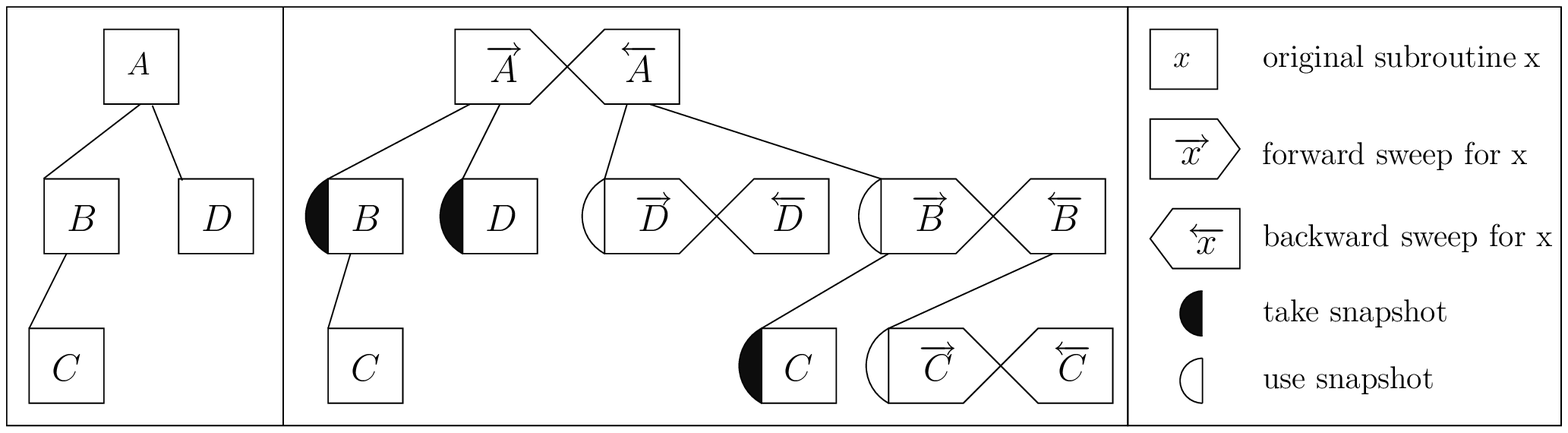}
\caption{Joint-All mode: Checkpointing all calls in Reverse Mode AD}
\label{fig:joint}
\end{figure}
Figure~\ref{fig:joint} shows (on the left) the call graph of an original program,
and the corresponding reverse-differentiated call graph, using the {\em Joint-All} mode,
where all procedure calls are checkpointed. This {\em Joint-All} mode is naturally the
basic mode, being the extreme trade-off that consumes time and saves memory.
Memory resources are finite, whereas execution time resources are not. Therefore this
choice is safest, especially if we assume that snapshots are generally smaller than the
corresponding tape.

\begin{figure}[!ht]
\begin{center}
\includegraphics[height=1.6in, width=2.3in]{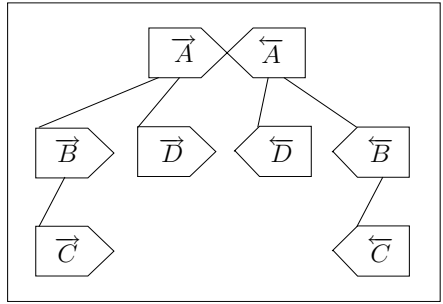}
\caption{Split-All mode: no Checkpointing in Reverse Mode AD}
\label{fig:split}
\end{center}
\end{figure}
Figure~\ref{fig:split} shows the other extreme alternative, which checkpoints no
procedure call. We call this alternative {\it Split-All} mode. In split mode 
the forward sweep and the backward sweep are implemented separately.
There is no duplicate execution, so no snapshot is required
and in theory the execution time is smallest.
On the other hand the peak size of the tape is highest.
Moreover, since the forward sweep and the backward sweep do not 
follow each other during execution, even the values of the local variables need to
be stored, which requires even more intermediate values in the tape.\\ 

\begin{figure}[!ht]
\centering
\includegraphics[height=1.6in, width=4.7in]{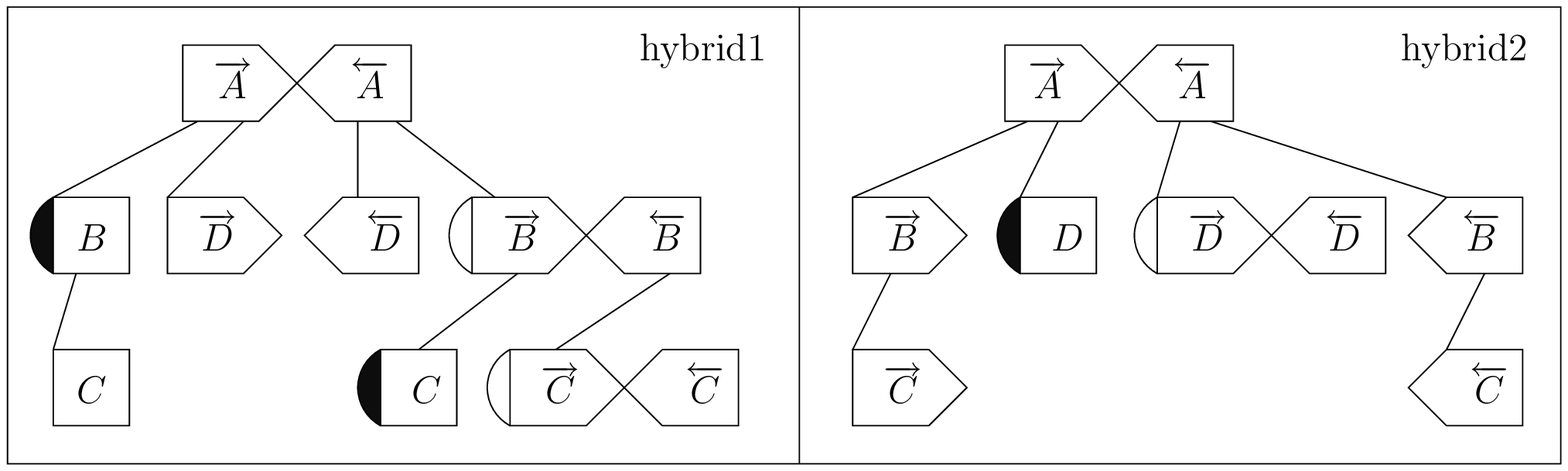}
\caption{Two hybrid approaches (split-joint)}
\label{fig:hybrid}
\end{figure}

Split-All and Joint-All modes are two extreme strategies. It is worth 
trying hybrid cases, we present a couple of cases in Figure~\ref{fig:hybrid}. 
The first strategy (hybrid1) implements the joint mode for all procedures except
for $D$. Conversely, the second strategy (hybrid2) implements the split 
mode for all procedures except for procedure $D$, which is checkpointed. \\

In order to have a more precise idea of the aforementioned trade-off we shall
simulate the performances of these four checkpointing strategies from
figures~\ref{fig:joint}, \ref{fig:split}, and~\ref{fig:hybrid},
for two motivating scenarios,
namely when {\it "tape $>$ snapshot"} and when {\it "tape $<$ snapshot"}.
We assume that all procedures require the same snapshot
and tape size. Also, we assume that each procedure has the same execution time.\\

For the first scenario, we set the memory size of the snapshot to 6 and
the memory size of the tape to 10. This setup corresponds to the usual assumption 
that the tape is bigger than the snapshot for procedures.
\begin{figure}[!ht]
\centering
\includegraphics[height=3in, width=5in]{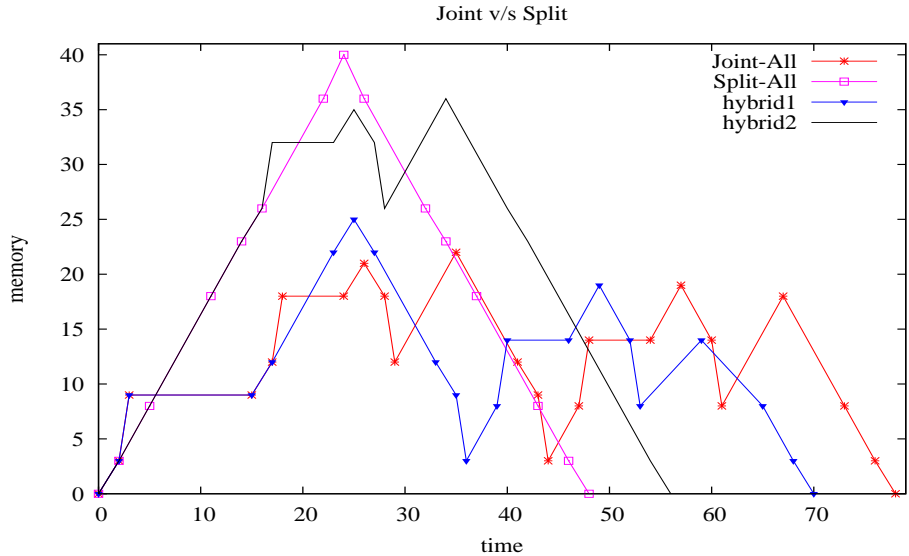}
\caption{\label{fig:comp} Numerical Simulation results, tape = 10, snapshot = 6}
\end{figure}
Figure~\ref{fig:comp} shows the behavior of the four checkpointing strategies.
As we expected, the curve that represents the joint
configuration shows the smallest memory use but the largest
execution time. Conversely, the curve that represents the split mode has
the highest peak of memory use but the shortest execution time.
Hybrid strategies range between these two extremes.\\

This scenario assumed that the tape is bigger than the snapshot. However, 
this a\-ssump\-tion is not always valid. Therefore we make a second simulation 
where we assume that the tape costs 6 in memory, and each snapshot costs 10.
\begin{figure}[!ht]
\centering
\includegraphics[height=3in, width=5in]{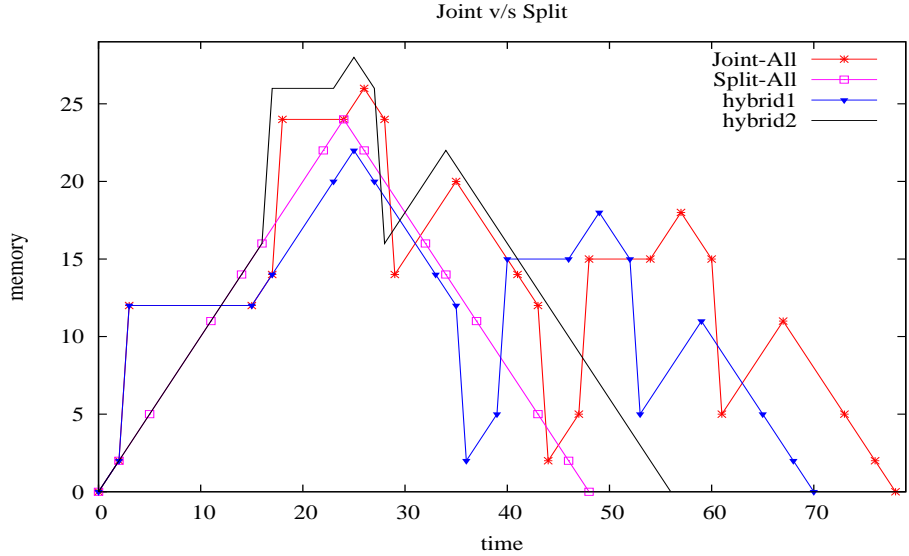}
\caption{\label{fig:comp2} Numerical Simulation results, tape = 6, snapshot = 10}
\end{figure}
Figure~\ref{fig:comp2} shows that Joint-All and Split-All modes are not
the extreme of the trade-off anymore. In fact, the extreme bounds in memory
consumption corresponds to the hybrid modes. We also notice that
the maximum peak of memory use is smaller than in the first simulation,
which is not surprising since it depends mostly on the tape size,
which is assumed smaller.
In this scenario, the 
advantage of checkpointing is less obvious because of the costs of snapshots,
therefore the Split-All mode is nearly the best in every respect.  \\

The real differentiated codes will have for every procedure different tape,
snapshot and execution time characteristics, making this motivating simulation look
a bit unreal. This gives us a feeling of the behavior of real
codes, but experiments with real code are mandatory.
Before we get to that, we shall briefly discuss the necessary implementation step.
\section{IMPLEMENTATION}\label{sec:imple}
We implemented the algorithms and data-flow analysis mentioned in the
previous section inside \textsc{tapenade} tool~\cite{Hasco04}, which is a source-to-source
AD engine. \textsc{tapenade} is written in JAVA and some modules are written in C. 
\textsc{tapenade} supports programs written in Fortran77 and Fortran90/95.
\subsection{Modifications of the Data-Flow Analyses}
The AD model that \textsc{tapenade} implements relies on several data-flow
analyses, all of them formally defined in~\cite{Hasco02, Hasco05}. However, these
analyses implicitly made the assumption of the Joint-All strategy.
The checkpointing strategy has s strong impact on adjoint liveness and TBR analyses,
which are interprocedural. More precisely, it impacts the way data-flow
information is propagated on the call graph during the bottom-up and top-down
analyses sweeps. \\

For example, since for a checkpointed segment the forward sweep is followed 
imme\-diately by the reverse sweep, we can use the fact that all original variables 
are useless at the end of the forward sweep. This is the foundation of the adjoint 
liveness analysis~\cite{Hasco05}. In the initial state of the AD tool where every 
call is checkpointed, this allowed the "adjoint-live" set at the tail of each 
procedure to be the empty set. The adjoint-liveness analysis can then proceed,
backwards inside the flow-graph of the procedure, progressively
accumulating variables into the set of live variables.
In the new situation where a procedure can be left in split mode, the initial
"adjoint-live" set at the tail of this procedure must change, and it depends of
the live variables in each of its calling sites. More precisely,
we shall set the live variables at the tail of a non-checkpointed procedure
(i.e. split mode) to the union of all the live variables just after
each of the call sites for this procedure. \\

In order to implement the mentioned adaptation we have to run the adjoint
liveness analysis twice. A first sweep runs bottom-up on the
original program call graph. In this sweep we build the effect of each
procedure on the set of live variables, to be used in each of its call sites.
The second run is top-down and accumulates the sets of live variable
after each call site, before it is used as
the initial set for the adjoint liveness analysis of every split 
procedure.\\ 

Similarly the TBR analysis had to be transformed. The TBR analysis runs
forward, from the head to the tail of each procedure. At the outer level
of the call graph, the analysis could run in only one bottom-up sweep.
Because TBR analysis now requires a context information in the case
of a non-checkpointed procedure, that will carry the union of the TBR
status just before the call sites, we had to add a
top-down sweep into the TBR analysis.

\subsection{General Implementation Notes}

Along with the modification of the analyses, the generation of the differentiated 
program must also be adapted. The AD model defined by equation~(\ref{form:call2})
shows that
the joint mode runs the backward sweep of $C$, $\overleftarrow{C}$,
immediately after its forward sweep $\overrightarrow{C}$.
When $C$ is a procedure, $\overrightarrow{C}$ and
$\overleftarrow{C}$ can be easily merged into a single procedure
$\overline{C}$. As a consequence, local variables of $C$ (and therefore of
$\overrightarrow{C}$) are still in scope when $\overleftarrow{C}$ starts,
and naturally preserve their values. This is no longer possible in split mode,
since procedure $\overrightarrow{C}$ and $\overleftarrow{C}$ are separated.
Consequently, local variables of $\overrightarrow{C}$ must be stored before
they vanish and restored when $\overleftarrow{C}$ starts.  
This was addressed in the implementation by adding an extension to the TBR analysis.
This extension looks for the locals variables that are necessary for the backward sweep,
when the end of the forward sweep is reached. These variables are
{\tt PUSH}'ed just at the end of the forward sweep and {\tt POP}'ed at 
the beginning of the backward sweep.  \\

We make the choice of generalization versus specialization, by allowing for only
one split mode per procedure. Even then, this requires care in naming
the procedures. We need to create up to four names (original, forward sweep, 
backward sweep and reverse differentiated) when split and joint strategies are 
combined. This problem is technical, but it has implications within the whole 
way \textsc{tapenade} handles the names of differentiated elements.\\

The split strategy is driven by the user by means of a directive ({\tt C\$AD
NOCHECKPOINT}) which is placed just before the procedure call, or through a 
command line option ({\tt -split "[list of procedure names]"}). 
The introduction of directives is a novel feature for \textsc{tapenade}.

\section{EXPERIMENTAL MEASUREMENTS}\label{sec:expe}

We applied the split mode to certain procedure calls, looking for 
experimental con\-fir\-mation of the intuitions from Section 5. In particular, 
we want to show the interest of letting the user drive the checkpointing strategy.\\

The procedures chosen to be split were the ones that best illustrate
the memory and run-time trade-off. The criteria to choose
procedures rely on two values, which can be obtained by studying the 
reverse generated code. These values are: the size of the snapshot and the size
of the tape. The implementation of both snapshot and tape is based on {\tt PUSH} calls, 
thus the measurements and comparisons between these values are straightforward.\\

\begin{figure}[!ht]
\centering
\includegraphics[height=2.5in, width=5in]{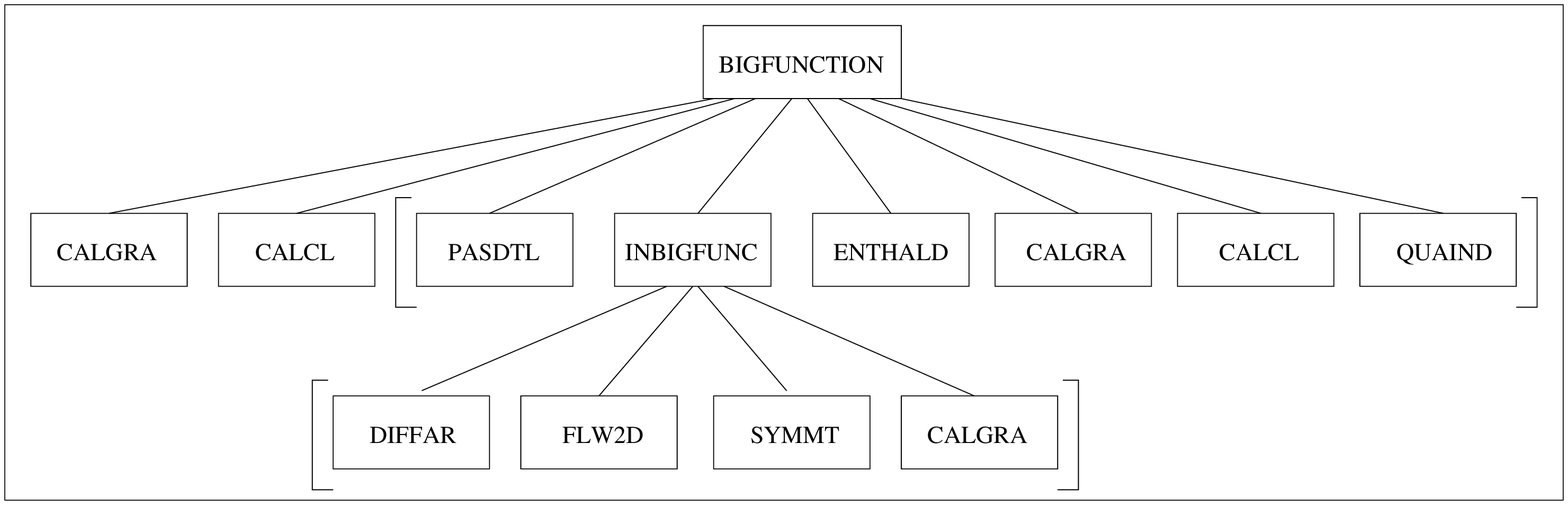}
\caption{\label{fig:uns2d} \textsc{uns2d} call graph.}
\end{figure}

\begin{figure}[!ht]
\centering
\includegraphics[height=3.0in, width=5.0in]{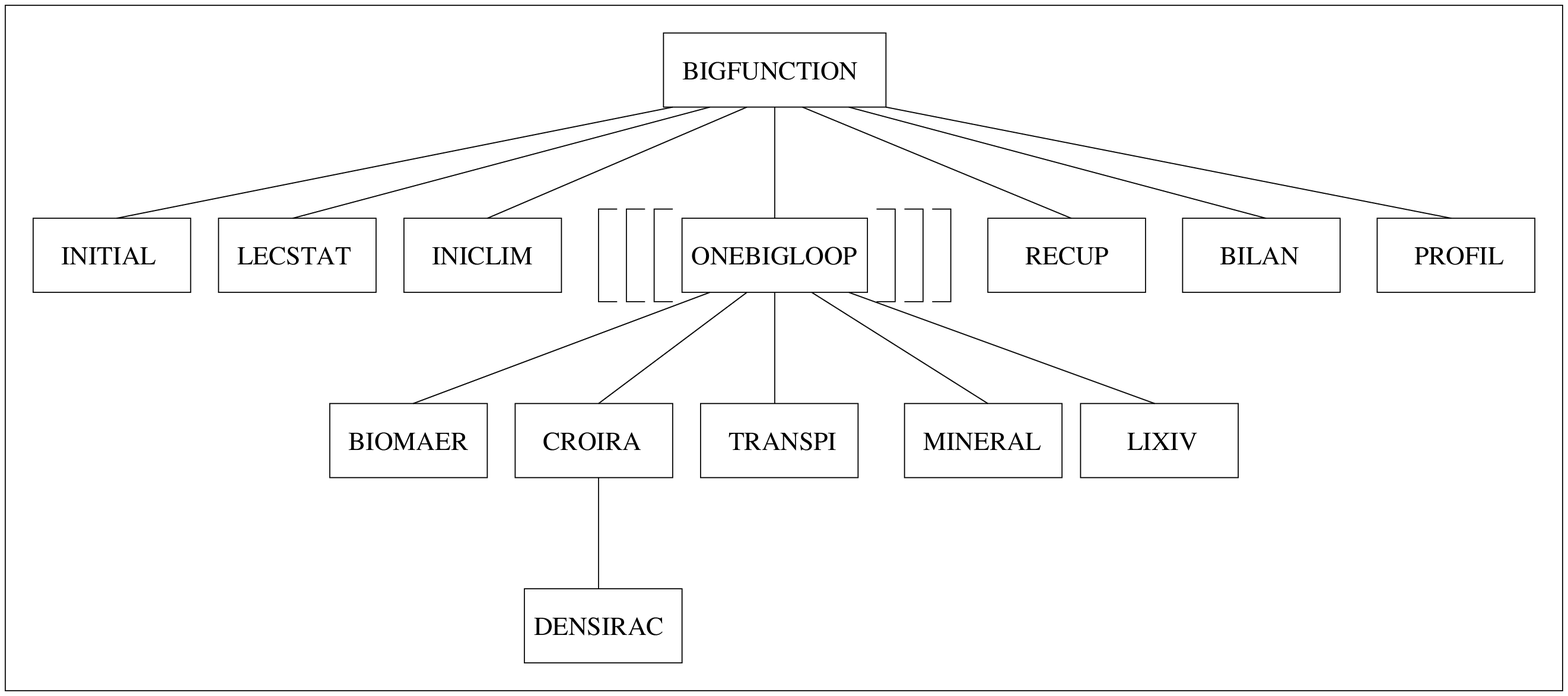}
\caption{\label{fig:call_stics} \textsc{stics} call graph.}
\end{figure}

In figures~\ref{fig:uns2d} and~\ref{fig:call_stics}, loops are denoted by 
square brackets. For instance, on Figure~\ref{fig:uns2d} we have two loops, 
one which involves from subroutine \textsc{pasdtl} to subroutine
\textsc{quaind}, and a second one which includes all \textsc{inbigfunc}'s 
procedures. In general, these loops are the segments of the programs that consume most 
of memory and time.\\
\subsection{Experiment I: UNS2D}
\textsc{uns2d} is a CFD solver. It has 2.055 lines of code (\textsc{loc}). The reverse
differentiated version has 2.200 \textsc{loc}. \\

\begin{Table}[!ht]
\centering
\begin{tabular}{|l|l||c|c||c|c|}
\hline
\multicolumn{2}{|c||}{Experiment}&\multicolumn{2}{c||}{Time}&\multicolumn{2}{c|}{Memory}\\
\hline
Id&Description &Total [s]&\textbf{\% gain}&Peak [Mb]&\textbf{\% gain}\\
\hline\hline
01&Joint-All strategy& 41.69&& 184.69&\\ \hline
02&split mode \textsc{calcl} (all call sites) & 37.66& 9.7& 167.53& 9.3\\ \hline
03&split mode \textsc{quaind} & 37.54& 9.9& 162.13& 12.2\\ \hline
04&split mode \textsc{calgra} (all call sites) & 36.63& 12.1& 163.92& 11.2\\ \hline
05&split mode \textsc{enthald} & 34.33& 17.6& 162.17& 12.2\\ \hline
06&split mode \textsc{inbigfunc} & 31.83& 23.6& 468.13& -153.5\\ \hline
07&02 and 05&33.95 &18.6 &163.20 &11.6 \\ \hline
08&03 and 06&31.75&23.8&446.82&-141.9\\ \hline
09&02, 04 and 05&35.81 &14.1 &174.45 &5.5 \\ \hline
10&02, 05 and 06& 35.49& 14.8& 533.23& -188.7\\ \hline
11&02, 03, 04 and 05& 38.50&7.6 &184.45 &0.13 \\ \hline
12&02, 04, 05 and 06& 30.92& 25.8& 408.88& -121.4\\ \hline
13&split mode all the above procedures& 32.67& 21.6& 443.56& -140.2\\ \hline
\end{tabular}
\caption{\label{table:uns2d} Memory and time performance for \textsc{uns2d}.}
\end{Table}

The first four experiments 02 - 05 of Table~\ref{table:uns2d} report
gain both in time and memory, reminding us of the case where $tape < snp$
(Figure~\ref{fig:comp2}). This is indeed what we observe when we measure the actual
sizes of tape and snapshot for the procedures in question. 
Therefore, when each of \textsc{calgra}, \textsc{calcl}, \textsc{quaind} or
\textsc{enthald} are split the program saves memory for the snapshot without
using as much for the tape. At the same time it saves time because the
procedure is not executed twice. \\

Experiment 06 exhibits a gain in time at the cost of a larger memory
use. As we suspected from the simulations on Figure~\ref{fig:comp}, this 
corresponds to the case where $snp < tape$. This confirms the intuition
that checkpointing is really worthwhile on large sections of program.
In this situation checkpointing  is really a time/memory trade-off. 
Therefore checkpointing \textsc{inbigfunc} (in other words the joint mode) is
a wise choice when memory size is limited.\\ 

Experiments 07 - 13 can be separated in two sets: whether \textsc{inbigfunc} is 
checkpointed (08, 10, 12 and 13) or not (07, 09 and 11). The separation criterion 
underlines the relative weight of the subroutine \textsc{inbigfunc}.\\ 

Experiments 07, 09 and 11 shows a remarkable behavior on the execution time
per\-for\-mance. We would expect the execution time savings of combined split
mode procedures to accumulate, as we observed in Figures~\ref{fig:comp}
and~\ref{fig:comp2}. Surprisingly, the execution time
for these experiments do not behave like that. In particular, the
experiment 11's execution time saving (3.18s) is smaller than the execution time 
savings (4.03s, 4.15s, 5.03s and 7.36s) for any of the procedures split individually.
We have at present no clear understanding of this behavior. It is likely that the
present model we have about the performances of checkpointed reverse programs, is
still insufficient to capture this behavior, and must be refined further.\\

As for concrete recommendations for this example,
we advise to apply split mode sparingly, only on one or two of subroutines
\textsc{calgra}, \textsc{calcl}, or \textsc{quaind} in the case where
there are strict memory constraints.
This allows for memory savings up to 12\%.
On the other hand, if memory is not an issue
and speed is, we recommend the configuration of experiment 12.
\subsection{Experiment II: SONICBOOM}
\textsc{sonicboom} is a part of a CFD solver which computes the residual of a
state equation. It has 14.263 \textsc{loc}, but only 818 \textsc{loc} to be differentiated, 
generating 2.987 \textsc{loc} of derivative procedures.  \\
\begin{figure}[!ht]
\centering
\includegraphics[height=1.7in, width=4.0in]{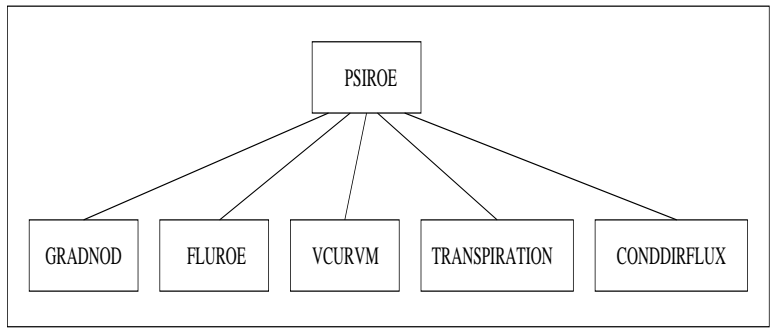}
\caption{\label{fig:call_alya} \textsc{sonicboom} call graph.}
\end{figure}

The first group of experiments 02 - 04 from Table~\ref{table:alya}, shows gains 
in execution time, because the procedures are executed only once. There is 
no gain in memory because the size of the snapshot and the tape are very close.\\

\begin{Table}[!ht]
\centering
\begin{tabular}{|l|l||c|c||c|c|}
\hline
\multicolumn{2}{|c||}{Experiment}&\multicolumn{2}{c||}{Time}&\multicolumn{2}{c|}{Memory}\\
\hline
Id&Description &Total [s]&\textbf{\% gain}&Peak [Mb]&\textbf{\% gain}\\
\hline\hline
01&Joint-All strategy& 0.2900&& 10.84&\\ \hline
02&split mode \textsc{vcurnvm}& 0.2725& 6.0& 10.84& 0.0\\ \hline
03&split mode \textsc{conddirflux}& 0.2699& 6.9& 10.84& 0.0\\ \hline
04&split mode \textsc{fluroe}& 0.2500& 13.8& 11.06& -2.0\\ \hline
05&split mode \textsc{gradnod}& 0.2374& 18.1& 18.77& -73.1\\ \hline
06&02 and 03&0.2624&9.5&10.84&0.0\\ \hline
07&04 and 05& 0.2374& 18.1& 19.00& -75.2\\ \hline
08&02, 03 and 04&0.2475&14.7&11.08&-2.2\\ \hline
09&02, 03 and 05&0.2360&18.6&18.77&-73.1\\ \hline
10&split mode all the above procedures& 0.2374& 18.1& 19.00& -75.2\\ \hline
\end{tabular}
\caption{\label{table:alya} Memory and time performance for \textsc{sonicboom}.}
\end{Table}

The experiments where \textsc{gradnod} is among the split subroutines exhibit the largest gain in
execution time. This is related to the fact that \textsc{gradnod} accounts for 
the largest part of the computation, and since the tape size grows like the
number of executed instructions, $tape(\textsc{gradnod})$ is much larger than
$snp(\textsc{gradnod})$. For the other procedures in this experiment we also have
$tape < snp$, but to a smaller extent. Therefore, everything behaves like in
the classical case of Figure~\ref{fig:comp}. In particular, there is no
procedure for which the split mode would give a gain in a memory
consumption.\\

It is worth noticing that the effect of the split mode is really an increase
in memory traffic rather than in memory peak size. For example splitting
\textsc{conddirflux} certainly results in a higher memory traffic, but the
local increase of the local memory peak is hidden by the main memory peak which
occurs just after $\overrightarrow{\textsc{gradnod}}$. We are currently
carrying new experiments and developing refined models that include this memory
traffic.\\

Practically for this experiment, our advice would be to run subroutines
\textsc{fluroe}, \textsc{vcurvm} and \textsc{conddirflux} (experiment 08) in 
split mode in any case, and this already gives a 14.7\% improvement in time at 
virtually no cost in memory. In the case where memory size is not
limited strongly, then it is advisable to run \textsc{gradnod} in split mode too, which 
gives an additional gain in time at the cost of a large increase in memory peak. 
\subsection{Experiment III: STICS}
\textsc{stics} is an agronomy modeling program. It has 21.010 \textsc{loc}, and 
the reverse differentiated code generated has 46.921 \textsc{loc}.
In the code of \textsc{stics}, we introduce three levels of nested loops around
subroutine \textsc{onebigloop} because this code simulates and unsteady process over 400
time steps. These nested loops are a manual modification that allow us to
perform checkpointing on various groups of time steps. We acknowledge that this
simplistic method is far from the known optimal strategy first described in~\cite{Griewank92}. \\
\begin{Table}[!ht]
\centering
\begin{tabular}{|l|l||c|c||c|c|}
\hline
\multicolumn{2}{|c||}{Experiment} &\multicolumn{2}{c||}{Time}&\multicolumn{2}{c|}{Memory}\\
\hline
Id&Description&Total [s]&\textbf{\% gain}&Peak [Mb]&\textbf{\% gain}\\
\hline\hline
01&Joint-All strategy &38.56&\textbf{}&229.23&\textbf{}\\ \hline
02&split mode \textsc{biomaer}&36.15&6.3&229.23&0.0\\ \hline
03&split mode \textsc{mineral}&35.78&7.2&229.28&0.0\\ \hline
04&split mode \textsc{densirac} &30.02&22.1&229.23&0.0\\ \hline
05&split mode \textsc{croira} &24.45&36.6&229.23&0.0\\ \hline
06&split mode \textsc{onebigloop} &23.75&38.4&229.75&-0.2\\ \hline
07&04 and 05&16.79&56.5&229.23&0.0\\ \hline
08&04 and 06&15.64&59.4&229.75&-0.2\\ \hline
08&05 and 06&11.71&69.6&206.81&9.8\\ \hline
09&04, 05 and 06 &3.93&89.8&149.11&34.9\\ \hline
09&03, 04, 05 and 06&3.92&89.8&149.11&34.9\\ \hline
09&split all the above procedures &3.90&89.9&149.11&34.9\\ \hline
\end{tabular}
\caption{\label{table:stics} Memory and time performance for \textsc{stics}.}
\end{Table}

For this experiment, the default (Split-All) strategy applied by
\textsc{tapenade} gave very bad results in time, with a slowdown factor of
about 100 from the original code to the reverse differentiated code. We made
some measurements of the tape sizes compared to the snapshot sizes, and we
found out that tape was much smaller than snapshot for subroutines
\textsc{densirac}, \textsc{croira} and \textsc{onebigloop}. This is a special
case of the situation of Figure~\ref{fig:comp2} and is reflected on the
experimental figures of Table~\ref{table:stics}. We see that split mode on
these three procedures gain execution time at no memory cost. Combined split
mode on the three procedures (experiment 09) gives an even better result.\\

The enormous gain in execution time makes the differentiated/original ratio
go down to about 7, which is what AD tools generally claim. In the \textsc{stics}
experiment, the exe\-cution time of the Split-All version did not come from the
duplicate exe\-cutions due to checkpointing but rather from the time needed to
{\tt PUSH} and {\tt POP} these very large snapshots. This suggests that a complete model
to study optimal checkpointing strategies should definitely take into account
the time spent for tape and snapshots operations.\\

Practically, in the \textsc{stics} example there is no doubt \textsc{densirac},
\textsc{croira} and \textsc{onebigloop} should be differentiated in split mode.
In addition, one can differentiate additional procedures in split mode, 
(e.g. \textsc{mineral}), but the additional execution time gain is marginal.  
\section{CONCLUSION, RELATED WORKS, FUTURE WORK}\label{sec:conclu}
This paper is a contribution towards the ultimate goal of optimally placing
checkpoints in adjoint codes built by reverse mode Automatic Differentiation.
We started from the observation that the strategy consisting in checkpointing
each and every procedure call is in general, although safe from the memory point of view,
far from optimal. Both simulations on very small examples,
and real experiments on real-life programs show that some procedures should
never be checkpointed, and that others may be checkpointed depending on the
available memory. The great variety of possible situations makes the objective
of automatic selection of checkpointing sites very distant. It seems therefore
reasonable to let the user drive this choice through an adapted user interface.
We discussed the developments that we made into the AD tool \textsc{tapenade}
to add this functionality. This new functionally allowed us to conduct
extensive experiments on real codes, that justified a posteriori our
hypotheses on this optimal checkpointing problem and suggest the relevant
criteria for a future helping tool namely, for each procedure, its execution
time, its tape and snapshot sizes, and the time required by tape {\tt PUSH} and 
{\tt POP} traffic.\\

Related works on optimal checkpointing have been conducted mostly on the model
case of loops of fixed-size iterations. Only in the particular sub-case where
the number of iterations in known in advance was an optimal scheme found
mathematically~\cite{Griewank92}. This gave rise to the
\textsc{treeverse}/\textsc{revolve}~\cite{Griewank99}
tool for an automatic application of this scheme.
In the case where the number of iterations is not known in advance,
a very interesting sub-optimal scheme was proposed in~\cite{HSRevolve05}.
We are not aware of optimal checkpointing schemes for
the case of an arbitrary call-tree or call graph. Notice that checkpointing
is not the only way to improve the performance of the reverse mode of AD. Local
optimization can reduce the computation cost of the derivatives by re-ordering
the sub-expressions inside derivatives~\cite{Griewank03}. Other optimizations implement a
fine-grain time/memory trade-off by storing expensive sub-expressions that
occur several times in the derivatives. In any case these are local
optimizations that only give a fixed small benefit. For large programs, only
nested checkpointing can make reverse differentiated codes actually run without
exceeding the memory capacity of the machine, and therefore the study of
optimal checkpointing schemes is an absolute necessity.\\

User-driven placement of checkpointing is an important step in this direction,
but further work is needed to help this placement or to propose a good enough 
automatic strategy. This could be based on execution time profiling of the
original program or even of the differentiated code itself. In any case, we
need to study the experimental figures found and to refine the model we have
built for the performance of reverse differentiated codes. In particular this
model must better take into account some of the surprising effects we have
found, such as time gains that do not add up. This suggests a process of
iterative improvements of the reverse differentiated codes, based on previous
runs, much like what is done in iterative compilation~\cite{Kisuki}.
\end{document}